\documentclass[sn-standardnature]{sn-jnl}
\makeatletter
\newcounter{biblast} 
\makeatother

\def\nat{Nature}
\def\aap{Astron.\ \& Astrophys.}

\def\apj{Astrophys.\ J.}

\def\apjl{Astrophys.\ J. Letters}
\def\apss{Ap\&SS}
\def\aj{Astron.\ J.}
\def\mnras{Mon.\ Not.\ R.\ Astron.\ Soc.}
\def\aapr{Astron.\ \& Astrophys.\ Rev.}
\def\pasp{Publ.\ Astron.\ Soc.\ Pac.}
\def\pasp{Publ.\ Astron.\ Soc.\ Pac.}
\def\solphys{Sol.~Phys.}

\def\icarus{Icarus}
\newcommand{\araa}{Ann. Rev. Astron. \& Astrophys.}
\newcommand{\ssr}{Space Science Reviews}

\newcommand{\jgr}{Journal of Geophysical Research (Space Physics)}
\newcommand{\stkm}{StKM\,1-1262}
\newcommand{\degr}{$^{\circ}$}

\jyear{2025}%

\begin{document}

\title[Radio Burst from a Stellar Coronal Mass Ejection]{Radio Burst from a Stellar Coronal Mass Ejection}


\author*[1,2]{\fnm{J.\,R.} \sur{Callingham}}\email{callingham@astron.nl}

\author*[3,4,5]{\fnm{C.} \sur{Tasse}}\email{cyril.tasse@obspm.fr}

\author[6,7]{\fnm{R.} \sur{Keers}} 

\author[1,2]{\fnm{R.\,D.} \sur{Kavanagh}} 

\author[1,8]{\fnm{H.} \sur{Vedantham}} 

\author[9,5]{\fnm{P.} \sur{Zarka}} 

\author[]{\fnm{S.} \sur{Bellotti}$^\text{10,11}$} 

\author[12]{\fnm{P. I.} \sur{Cristofari}} 

\author[1,8]{\fnm{S.} \sur{Bloot}} 

\author[1,8]{\fnm{D. C.} \sur{Konijn}} 

\author[13]{\fnm{M. J.} \sur{Hardcastle}} 

\author[9,5,14]{\fnm{L.} \sur{Lamy}} 

\author[12,15]{\fnm{E. K.} \sur{Pass}} 

\author[16]{\fnm{B. J. S.} \sur{Pope}} 

\author[]{\fnm{H.} \sur{Reid}$^\text{17}$} 

\author[]{\fnm{H. J. A.} \sur{R\"{o}ttgering}$^\text{10}$}

\author[1,10]{\fnm{T. W.} \sur{Shimwell}} 

\author[1]{\fnm{P.} \sur{Zucca}} 

\affil[1]{\orgdiv{ASTRON}, \orgname{Netherlands Institute for Radio Astronomy}, \orgaddress{\street{Oude Hoogeveensedijk 4}, \city{Dwingeloo}, \postcode{7991\,PD}, \country{The Netherlands}}}

\affil[2]{\orgdiv{Anton Pannekoek Institute for Astronomy}, \orgname{University of Amsterdam}, \orgaddress{\street{Science Park 904}, \postcode{1098\,XH}, \city{Amsterdam}, \country{The Netherlands}}}

\affil[3]{\orgdiv{LUX}, \orgname{Observatoire de Paris, Université PSL, CNRS}, \orgaddress{\city{Meudon}, \postcode{792190}, \country{France}}}

\affil[4]{\orgdiv{Centre for Radio Astronomy Techniques and Technologies (RATT)}, \orgname{Department of Physics and Electronics, Rhodes University}, \orgaddress{ \city{Makhanda}, \postcode{6140}, \country{South Africa}}}

\affil[5]{\orgdiv{Observatoire Radioastronomique de Nançay (ORN)}, \orgname{Observatoire de Paris, CNRS, PSL, Université d'Orléans, OSUC}, \orgaddress{\street{route de Souesmes}, \city{Nançay}, \postcode{18330}, \country{France}}}

\affil[6]{\orgname{Max Planck Institute for Solar System Research}, \orgaddress{\street{Justus-von-Liebig-Weg 3}, \postcode{37077}, \city{Göttingen}, \country{Germany}}}

\affil[7]{\orgdiv{Faculty of Electrical Engineering, Information Technology, and Physics}, \orgname{Technische Universität Braunschweig}, \orgaddress{\street{Hans-Sommer-Strasse 66}, \postcode{38106}, \city{Braunschweig}, \country{Germany}}}

\affil[8]{Kapteyn Astronomical Institute, University of Groningen, PO Box 72, 97200 AB, Groningen, The Netherlands}

\affil[9]{\orgdiv{LIRA}, \orgname{Observatoire de Paris, Université PSL, CNRS, Sorbonne Université, Université de Paris}, \orgaddress{\street{5 place Jules Janssen}, \city{Meudon}, \postcode{92195}, \country{France}}}

\affil[10]{\orgdiv{Leiden Observatory}, \orgname{Leiden University}, \orgaddress{\street{PO\,Box 9513}, \postcode{2300\,RA}, \city{Leiden}, \country{The Netherlands}}}

\affil[11]{\orgdiv{Institut de Recherche en Astrophysique et Plan\'etologie}, \orgname{Universit\'e de Toulouse, CNRS}, \orgaddress{\street{IRAP/UMR 5277, 14 avenue Edouard Belin}, \postcode{31400}, \city{Toulouse}, \country{France}}}

\affil[12]{\orgdiv{Center for Astrophysics}, \orgname{Harvard \& Smithsonian}, \orgaddress{\street{60 Garden Street}, \postcode{02138}, \city{Cambridge, MA}, \country{United States of America}}}

\affil[13]{\orgdiv{Centre for Astrophysics Research}, \orgname{University of Hertfordshire}, \orgaddress{\street{College Lane}, \postcode{AL10 9AB}, \city{Hatfield}, \country{United Kingdom}}}

\affil[14]{\orgdiv{Kavli Institute for Astrophysics and Space Research}, \orgname{Massachusetts Institute of Technology}, \orgaddress{ \postcode{02139}, \city{Cambridge, MA}, \country{United States of America}}}

\affil[15]{\orgdiv{LAM}, \orgname{Aix-Marseille Universit\'{e}, CNRS, CNES}, \orgaddress{\street{38 Rue Fr\'{e}d\'{e}ric Joliot Curie}, \postcode{13013}, \city{Marseille}, \country{France}}}

\affil[16]{\orgdiv{Department of Physics and Astronomy}, \orgname{Maquarie University}, \orgaddress{\postcode{NSW 2109}, \city{Sydney}, \country{Australia}}}

\affil[17]{\orgdiv{Mullard Space Science Laboratory}, \orgname{University College London}, \orgaddress{\street{Holmbury, Hill Rd}, \postcode{RH5 6NT}, \city{Dorking}, \country{United Kingdom}}}
            

\abstract{Coronal mass ejections (CMEs) are massive expulsions of magnetised plasma from a star, and are the largest contributors to space weather in the Solar System \citep{2015A&A...580A..80K,2016Ap&SS.361..253B}. CMEs are theorized to play a key role in planetary atmospheric erosion, especially for planets that are close to their host star \citep{2007AsBio...7..167K,2016ApJ...826..195K,2022SpWea..2003164V}. However, such a conclusion remains controversial as there has not been an unambiguous detection of a CME from a star outside of our Sun. Previous stellar CME studies have only inferred the presence of a CME through the detection of other types of stellar eruptive events \citep{1990A&A...238..249H,2016A&A...590A..11V,1999A&A...350..900F,2022ApJ...936..170L}. A signature of a fast CME is a Type\,II radio burst \citep{gopalswamy2006,gopalswamy2008}, which is emitted from the shock wave produced as the CME travels through the stellar corona into interplanetary space. Here we report an analogue to a Type\,II burst from the early M\,dwarf StKM\,1-1262. The burst exhibits identical frequency, time, and polarisation properties to fundamental plasma emission from a solar Type\,II burst. We demonstrate the rate of such events with similar radio luminosity from M dwarfs are 0.84$^{+1.94}_{-0.69} \times$10$^{-3}$ per day per star. Our detection implies that we are no longer restricted to extrapolating the solar CME kinematics and rates to other stars, allowing us to establish the first observational limits on the impact of CMEs on exoplanets.} 

\maketitle
\backmatter

Coronal Mass Ejections (CMEs) from the Sun play a fundamental role in shaping the space weather of the Solar System, from driving aurorae to eroding planetary atmospheres \citep{2015A&A...580A..80K,2016Ap&SS.361..253B}. Since the discovery of exoplanets, significant effort has turned to determining the rate and intensities of CMEs around other stars as the persistent impact of CMEs has the potential to strip a planet of its atmosphere \citep{2007AsBio...7..167K,2016ApJ...826..195K,2022SpWea..2003164V}. We know that younger solar analogues and M\,dwarfs have much higher flare activity levels than the Sun \citep{Feinstein2020,Gunther2019}, but it is difficult to know what effect the radically different stellar coronal and magnetospheric conditions will have on the production of CMEs. Understanding CMEs from M\,dwarfs is particularly pertinent as they host the largest fraction of conventional habitable-zone exoplanets \citep{2015ApJ...807...45D}. Since the conventional habitable zones around M\,dwarfs are substantially closer to the star than in the Solar System \citep{1993Icar..101..108K}, the planets around M dwarfs could be subjected to more frequent, higher energy CME impacts than those experienced by Earth \citep{2007AsBio...7..167K,2016ApJ...826..195K}.

Unfortunately, CMEs are so dim that we have been hitherto restricted to using indirect methods to infer their presence. For example, promising evidence of stellar CMEs includes blueshifts of chromospheric lines \citep{1990A&A...238..249H,2016A&A...590A..11V,2022NatAs...6..241N,2025arXiv251022110N} and extreme UV/X-ray coronal dimming \citep{1999A&A...350..900F,2019NatAs...3..742A,veronig21,2022ApJ...936..170L}. However, such Doppler shift and dimming measurements do not inform us if the fast-moving plasma was still retained in the stellar magnetosphere -- implying that the plasma may never impact a putative planet. Considering that the magnetic fields of M\,dwarfs can be over three orders of magnitude stronger than the global solar magnetic field \citep{Alvarado-Gomez2018ApJ...862...93A}, we require a tracer that establishes that plasma has been causally disconnected from the stellar magnetosphere. 

One such tracer that reveals that mass has been ejected from a stellar magnetosphere into interplanetary space is a Type\,II radio burst \citep{paynescott1947,wild1950}. A Type\,II burst is commonly emitted when a CME is super-Alfv\'{e}nic, implying that a shock has been produced that causally disconnects the plasma from the stellar magnetosphere. The shock excites Langmuir waves as it propagates out of the solar atmosphere, producing coherent radio emission \citep{1985ARA&A..23..169D}. Type\,II radio bursts typically last a few minutes, with the emission lanes drifting from high to low frequencies over time, encoding the change in plasma density as the shock propagates. Type\,II bursts are observed to occur often at both fundamental and harmonic plasma emission frequencies, with the fundamental emission able to reach high brightness temperatures and circular-polarisation fractions \citep{1985ARA&A..23..169D}. Solar radio observations have demonstrated that there is a strong ($\gtrsim80\%$) association of Type\,II bursts with the fastest ($\gtrsim 1800$\,km\,s$^{-1}$) CMEs \citep{gopalswamy2006,gopalswamy2008}. While there been rare instances where solar Type\,II bursts are not associated with CMEs \citep{2015ApJ...804...88S}, such CMEs have relative slow speeds.

While a Type\,II burst is unambiguous evidence of plasma escaping from a stellar magnetosphere, there have been no firm detections of Type\,II bursts from other stars despite significant observational effort \citep{Crosley2016ApJ...830...24C, Crosley2018ApJ...862..113C,2017IAUS..328..243O, Villadsen2019ApJ...871..214V,Zic2020}. The previous non-detections of stellar Type\,II bursts may be due to sensitivity limitations, the short duration of observations, magnetic confinement of CMEs on radio-bright stars \cite{Alvarado-Gomez2018ApJ...862...93A}, or a high Alfv\'en speed that prevents shock formation in other stellar magnetospheres \cite{Alvarado-Gomez2020ApJ...895...47A}. Instead, radio observations of M~dwarfs have found other types of coherent radio bursts, including hours-long events attributed to the electron-cyclotron maser instability (ECMI) \cite{Villadsen2019ApJ...871..214V,Zic2019MNRAS.488..559Z,Callingham_crdra,Bastian22} that often do not have a direct solar analogue. 

Considering that the radio emission of solar Type\,II bursts is stochastic and occurs predominantly at low frequencies ($\lesssim 300$\,MHz), low-frequency wide-field surveys with high sensitivity on timescales of minutes are potentially a powerful tool for discovering a stellar Type\,II burst. With this aim, we have conducted the LOw-Frequency ARray (LOFAR) Two-metre Sky Survey \citep[LoTSS;][]{2022A&A...659A...1S}, covering the Northern sky at sensitivities $\lesssim 2$\,mJy per minute. Each $\approx16$sq. deg. field was targeted for approximately 8\,hours, resulting in $\sim$86,000 stars within 100\,pc being observed and searched (Tasse et al., accepted).

\begin{figure}[h]
    \centering
    \includegraphics[width = \textwidth]{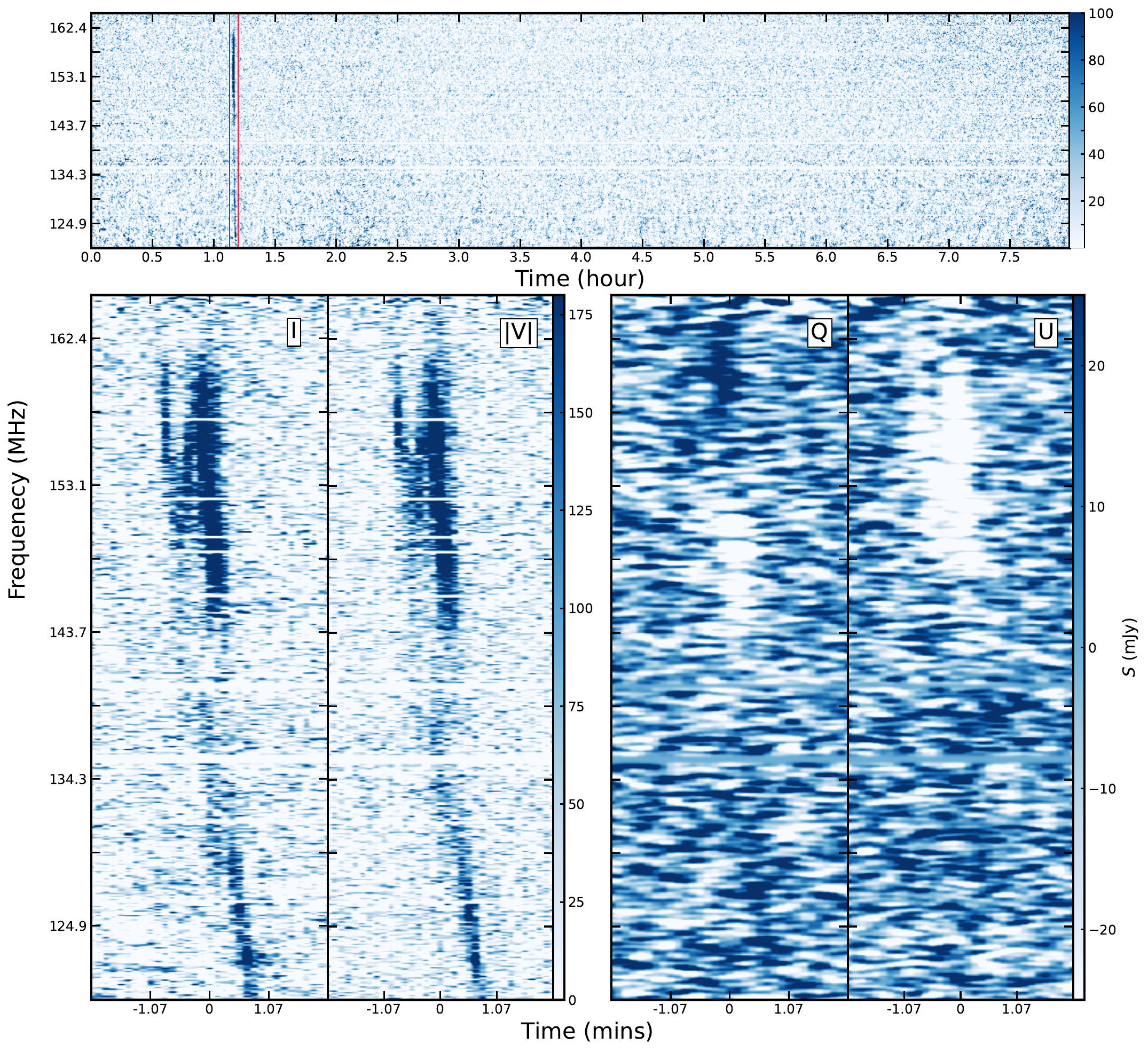}
    \caption{\textbf{Dynamic spectra of the burst for different polarisations and durations.} The total intensity dynamic spectrum for the entire 8\,h observation is shown in the top panel, with the burst bracketed by two red lines. The burst is centered in the bottom panels, with Stokes I, absolute V, Q, and U shown left to right, respectively. The dynamic spectra of Stokes Q and U are smoothed with a Gaussian filter with a kernel of 1 pixel to enhance the visbility of the fainter emission. Note that the colour scale of Stokes I and V are different to that of Stokes Q and U, as demonstrated by distinct colour bars.}
    \label{fig:dyn_spec_all}
\end{figure}

\begin{figure}[h]
    \centering
    \includegraphics[width = \textwidth]{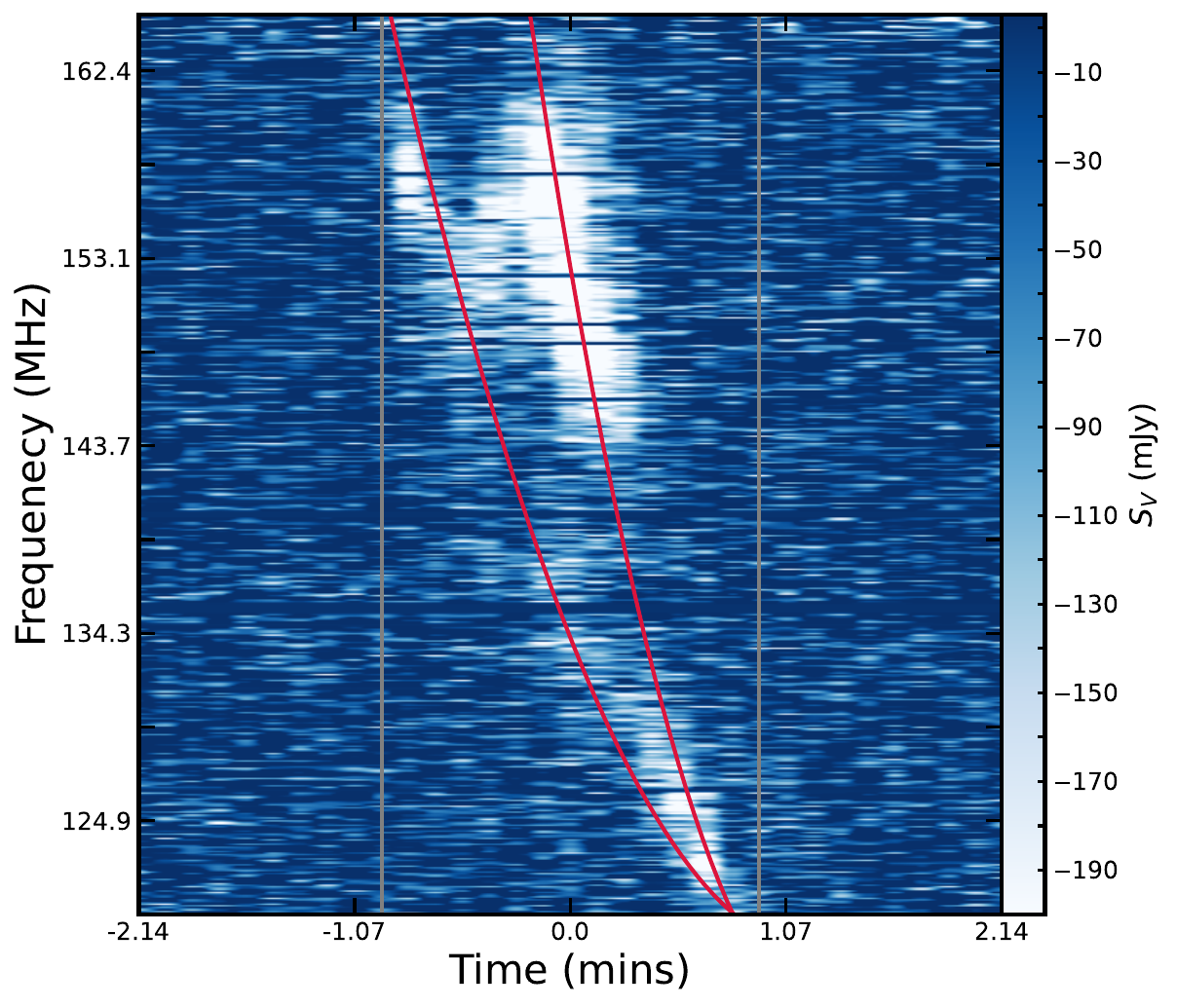}
    \caption{\textbf{A ridge-crawler fit to the two different emission lanes evident in the circularly polarised burst.} The red lines trace the two fits, and the vertical grey lines demarcate the area over which the fit was searched. The two fits intersect at the lowest frequency channel since that channel contains the largest signal-to-noise at the edge of the band. This is likely a product of the limited bandwidth of our observation.}
    \label{fig:dyn_spec_fit}
\end{figure}

During this survey, we detected a stellar swept-frequency radio burst that displays a striking similarity to a solar Type\,II burst. As shown in Figure\,\ref{fig:dyn_spec_all}, the burst lasts for $\approx$2\,mins, sweeping from 166 to 120\,MHz. This corresponds to a median drift rate of $\approx\,-0.62\pm$0.22\,MHz\,s$^{-1}$, as fit in Figure\,\ref{fig:dyn_spec_fit}. The burst was detected $\approx$1.2 hours into the LOFAR observation, with no other emission detected in the remaining $\approx$6.8 hours of observation. It has an average flux density of $\approx$110\,mJy, but reaches a peak flux density of 440$\pm$20\,mJy. The burst also appears to display two different lanes of emission, as highlighted in Figure\,\ref{fig:dyn_spec_fit}, which could be interpreted as band-splitting. Quadratic fits to the two sub-bands are shown in Figure\,\ref{fig:dyn_spec_fit}, with the emission between the bands potentially analogous to herringbone structures. The burst is highly circularly polarised at an average of $\approx$90\%, and displays a small amount of linear polarisation for part of its duration ($\approx$10\%). 

As shown in Figure\,\ref{fig:circ_pol_image}, the burst is localised to the active M0V star \stkm. The star is located at 40.8\,pc, is not a known binary, has a 1.241$\pm$0.003\,d optical photometric rotation signal \citep{2024AJ....167..189C}, an effective temperature of 3916$\pm$7\,K, and a quiescent soft (0.2-2.0\,keV) X-ray luminosity of 3.2$\pm$0.1 $\times$10$^{29}$ erg\,s$^{-1}$ (determined from a \emph{XMM-Newton} observation; see Methods). From Zeeman-Doppler imaging, the global magnetic field of \stkm~has a poloidal-dipolar, non-axisymmetric topology with an average magnetic field strength of $\approx300$\,G, and a $v\sin i$ of 25.9$\pm$ 0.2\,km\,s$^{-1}$ (Bellotti et al., accepted). Assuming a stellar radius for an M0V star of 0.63$\pm$0.01\,$R_{\odot}$ \citep{2019ApJ...871...63M}, the combination of $v\sin i$ and the measured photometric period implies we are viewing the star equator-on. 

With these known characteristics of \stkm, we can test if the detected burst fits within the solar paradigm of Type\,II bursts. For example, the velocity of the plasma ejected can be estimated given the drift rate of the burst, coronal scale height of the star, and frequency of emission. As the burst propagates through the stellar corona, the frequency $\nu$ of radio emission is expected to vary in time $t$ as \citep{Crosley2016ApJ...830...24C}

\begin{equation}\label{eqn:velocity}
    \frac{d\nu}{dt}=\frac{\partial\nu}{\partial n_e}\frac{\partial n_e}{\partial h}\frac{\partial h}{\partial s}\frac{\partial s}{\partial t},
\end{equation}

\noindent where $n_{e}$ is the electron density, $h$ is radial height above the star, and $s$ is the distance traveled by the shock. For a barometric atmosphere, negligible acceleration, and a shock traveling parallel to our line-of-sight, Equation\,\ref{eqn:velocity} can be reduced to $\frac{d\nu}{dt} = -\nu v_{s} / 2H$, where $v_s$ is the speed of the shock and $H$ the density scale height. The density scale height of \stkm~can be estimated from its X-ray luminosity (see Methods). For the detected burst, we derive a shock velocity at 144\,MHz of 2400$\pm$600\,km\,s$^{-1}$.

To place this shock velocity into the solar context, $\gtrsim 95\%$ solar CMEs with velocities $\ge$2400\,km\,s$^{-1}$ are accompanied by a Type\,II burst \citep{2008ApJ...674..560G,gopalswamy2008,2023A&A...675A.102K}. Such a tight correlation is likely because plasma with a velocity exceeding 2400\,km\,s$^{-1}$ is almost always super-Alfv\'{e}nic, regardless of the local density and magnetic field inhomogeneities in the solar corona \citep{2008ApJ...674..560G,gopalswamy2008}. Furthermore, only $\approx$0.05\% of CMEs occur with velocities as high as $\ge$2400\,km\,s$^{-1}$ on the Sun \citep{2009EM&P..104..295G}, and the average CME rate over the solar cycle is $\sim$1.2 per day \citep{2009EM&P..104..295G}. If we assume that our sensitivity horizon to Type\,II bursts is the distance to \stkm, we have surveyed 5,034 main sequence stars (3,555 stars of spectral type $\sim$M0 to M6) for 8 hours each in our LOFAR survey (Tasse et al., accepted). Therefore, we could expect $\sim$1 burst like that shown in Figure\,\ref{fig:dyn_spec_all}, with the same radio luminosity as that observed for Type\,II bursts on the Sun \citep{2023A&A...675A.102K}. This also implies that the rate of such events with similar radio luminosity from M0 to M6 stars are $0.84^{+1.94}_{-0.69} \times$10$^{-3}$ per day per star (1-$\sigma$ uncertainty). Therefore, while the drift rate of the burst and derived velocity are fast, solar Type\,II bursts have been observed with similar properties \citep{2008ApJ...674..560G,gopalswamy2008}. Finally, while we note there are exceptional solar Type\,II bursts that do not necessarily accompany a CME, in almost all cases this still requires the expulsion of plasma from the magnetosphere of the Sun \citep{2021ApJ...909....2M}.

While the velocity and morphology of the burst fits within a solar paradigm for Type\,II bursts, it has other non-standard features. These non-standard features can be explained within the solar Type\,II emission framework. For example, solar Type\,II radio bursts are characterised by their emission at both fundamental and harmonic frequencies. Since the circular polarisation fraction of our burst exceeds 60\%, it implies that we are detecting emission at the fundamental frequency as only the ordinary mode of the plasma can escape \citep{1985ARA&A..23..169D}. We do not observe the harmonic emission at later times (top panel, Figure\,\ref{fig:dyn_spec_all}). Solar observations demonstrate that it is possible for the harmonic emission never to extend down to as low frequency as the fundamental emission, due to differences in propagation effects, locations of the emission sites, and the group velocities of the electromagnetic waves in the solar corona \citep{2023A&A...675A.102K}. 

Additionally, linear polarisation is canonically expected to be negligible in solar radio bursts due to the high amount of Faraday rotation experienced by the emission as it traverses the corona \citep{1973SoPh...29..149G}. However, recent work has challenged such assumptions, with linear polarisation fractions of $\approx$25\% detected from solar Type\,II bursts \citep{2025ApJ...988L..73D}. The high circular polarisation fraction of the burst is also large for a standard solar Type\,II burst, but fundamental plasma emission is expected to be generated close to 100\% circularly polarised \citep{1985ARA&A..23..169D}. Fundamental emission from solar Type\,II bursts are often observed with a low polarisation fractions due to scattering and depolarisation of the emission as it crosses overlying plasma layers in the corona \citep{1966AuJPh..19..209S}. Therefore, we interpret the preserved high degree of polarisation of our burst as indicative of a relatively clean (i.e. low plasma density and stable magnetic field structure) sight-line to the plasma emission site. Such low-density, stable magnetospheric structures have previously been proposed to exist in M dwarfs due to their strong magnetic fields \citep{Villadsen2019ApJ...871..214V,Zic2020,Callingham_crdra}. 

Finally, the distance and flux density of the detected burst implies a high brightness temperature for the emission. To derive a physical emitting region, we assume the radio emission occurs at $\sim3$ times the stellar radius, calculated by applying a fourfold Newkirk density model for the corona of \stkm~with a base density of $\sim 0.6 \times 10^{10}$\,cm$^{-3}$ \citep{1995A&A...295..775M,2004ApJ...617..508T,2023A&A...675A.102K}. Assuming the angular spread of the emission  region $\Delta\psi$ is identical to that observed on the Sun \citep{2018ApJ...856...39C}, namely $\Delta\psi = 13^{\circ}(r/R_{*})^{0.22}$, we find the emitting region is approximately 55\% of the photospheric surface of \stkm. Such an emitting region would produce a brightness temperature of $\approx1.5 \times 10^{15}$\,K.

Fundamental emission can reach such brightness temperatures, provided the injected hot plasma is $\sim$40 times hotter than the ambient coronal temperature of \stkm~(see Methods). Such temperatures have been measured for flaring M dwarfs \citep{Robrade2010}. While the ECMI mechanism can recover some of the characteristics of the burst, such as the high circularly polarised fraction and brightness temperature, it struggles to generate the observed frequency sweep unless contrived magnetic geometry and viewing angles are assumed (see Methods). Furthermore, ECMI expects \stkm~to emit periodically, modulated by the rotation of the star. Follow-up LOFAR observations that have covered the full rotation period of \stkm~have failed to detect the star again. Finally, the burst would be the most luminous Type\,II burst detected on the Sun by approximately four-orders of magnitudes. However, the rate we determine in the LoTSS survey at which such luminous bursts occur is broadly consistent with the solar Type\,II burst rate \citep{2017SpWea..15.1511G}.

Assuming the burst is a stellar Type\,II burst, we can use its emission characteristics to measure the properties of the ejected plasma from \stkm. The frequency of the emission of the burst is determined by the electron density $n_{e}$\,[cm$^{-3}$] found in the background plasma, namely $\nu = 9\times10^{-3} \sqrt{n_{e}}$~MHz. The frequency of the emission implies that the density of the plasma ejected is greater than $\sim 3 \times 10^{8}$\,cm$^{-3}$. Assuming the corona of \stkm~is in hydrostatic equilibrium, the plasma was deposited at $\sim$3 times the stellar radius $R_{*}$ (assuming a base density of $\sim 0.6 \times 10^{10}$\,cm$^{-3}$, which is common for active M dwarfs \citep{2004ApJ...617..508T,2023A&A...675A.102K}). At such a distance, the deposited electron density is an order of magnitude larger than what is commonly simulated for CME impacts on exoplanets \citep{2022SpWea..2003164V,2016ApJ...826..195K}. Furthermore, since such radio emission only occurs because the shock is super-Alfv\'{e}nic, that allows us to put an upper-limit of 19\,G for magnetic field of the corona at a distance of $\sim$3 $R_{*}$. 

While the magnetic field of the star will determine how such plasma will evolve, if we assume a Parker model and CME mass scaling (total mass $\gtrsim$10$^{15}$\,g \citep{1992ApJ...390L..37G}), such a CME will produce a ram pressure of $\gtrsim$0.1\,dyn\,cm$^{-2}$ at 0.2\,AU from \stkm, the inner boundary of the conventional habitable zone for a M0V star \citep{1993Icar..101..108K}. Such pressure would be capable of compressing planetary magnetospheres to their surface for even a strong terrestrial magnetic field strength of $\sim$3\,G \citep{2016ApJ...826..195K}. 

We have demonstrated that we have detected a radio burst from an M dwarf that has the characteristics of a stellar Type\,II burst, allowing us to: 1) unambiguously conclude that there has been hot plasma released from \stkm~into its interplanetary medium; and, 2) calculate the density of the plasma that was released, implying a devastating impact for a putative exoplanet in the conventional habitable zone of \stkm. Our work means that our empirical benchmarks for the rate and intensities of stellar CMEs are no longer solely determined by the Sun scaled to vastly different stellar magnetic field topologies, ages, strengths, and coronal densities. Our work also supports the rarity of detectable stellar CMEs from M dwarfs, as also suggested by other studies \citep{Crosley2018ApJ...862..113C}. The detection of such a burst demonstrates we are opening the field of stellar CME studies with LOFAR and the upcoming Square Kilometre Array.

\newpage 
\newpage
\newpage
\begin{figure}[h]
    \centering
    \includegraphics[width = 0.5\textwidth]{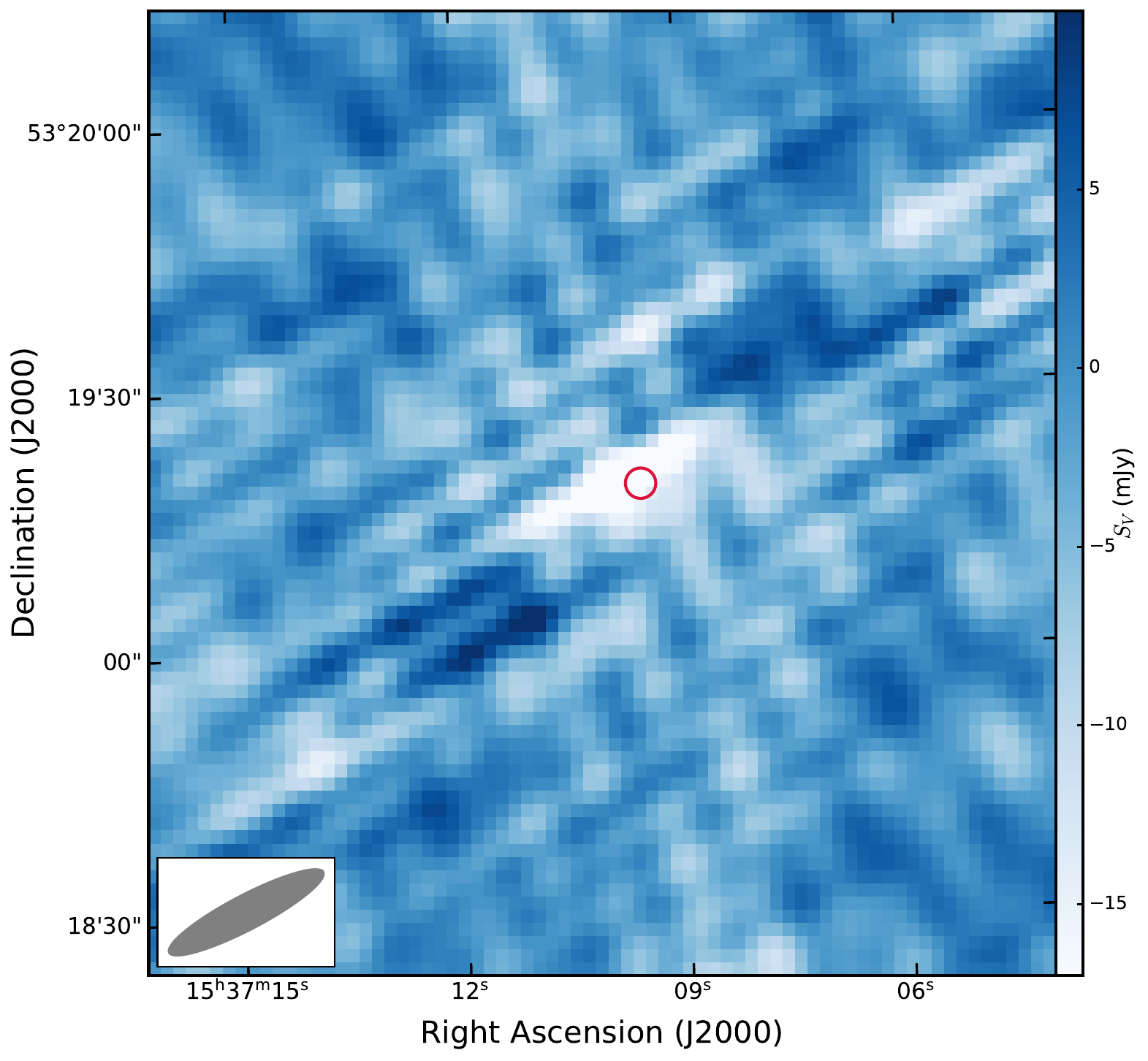}
    \caption{\textbf{Circularly polarised dirty map of \stkm, imaged over the duration of the $\approx$2 minute burst.} The red circle identifies the \emph{Gaia}\,Data Release 3 \citep{2023A&A...674A...1G} position of \stkm~at the time of the LOFAR observation and has a radius of 2$''$, corresponding to the astrometric uncertainty of the radio position. The size and shape of the instantaneous synthesised beam is shown in the bottom left corner.}
    \label{fig:circ_pol_image}
\end{figure}
\newpage

\section*{Methods}

\subsection*{Radio data acquisition and reduction}

The \stkm~burst discussed here (see Figure\,\ref{fig:dyn_spec_all}) was detected as one of the brightest bursts in a radio survey monitoring a predefined sample of $\sim200,000$ nearby stars using LOFAR \citep{vanHaarlem2013} data at $\sim150$ MHz. Details of the data processing are given in Tasse et al. (accepted), but we give a brief overview here.
 
The LOFAR Two Meter Sky Survey (LoTSS) \citep{Shimwell17,Shimwell22},
initiated in 2016, aims to produce deep, high-resolution images of the
northern hemisphere by observing in the $\sim120-165$\,MHz frequency
range. Upon completion, it will include approximately 3,000 pointings,
each involving 8-12 hours of integration, resulting in $\sim340$ PB of
raw data. To manage this volume, the data is flagged for terrestrial
radio frequency interference, averaged, and compressed by a factor of
$\sim17$. Producing images with LoTSS is challenging due to the large
$\sim5$ degree field of view of individual observations, the complexity
of antenna primary beams, ionospheric disturbances, and the big data
rates. To overcome these obstacles, we developed a method that
implements an efficient, wide-field adaptive optics system
generalized for polarization \citep{Tasse14b, Smirnov15, Tasse18}. The
LoTSS pipeline is based in these findings, and operates using the
ddf-pipeline\footnote{\url{https://github.com/mhardcastle/ddf-pipeline}}
framework. By alternating between
kMS\footnote{\url{https://github.com/saopicc/killMS}} and
DDFacet\footnote{\url{https://github.com/saopicc/DDFacet}} softwares,
ddf-pipeline iteratively solves the direction-dependent calibration
and imaging problems, and produces thermal noise-limited,
high-fidelity images, in a totally automated way \citep{Tasse21}.

In order to search for stellar variability within the observation
scans, we introduced an additional processing step to the pipeline
that consists in coherently averaging the residual visibilities toward
selected stellar sources. For a given direction in the sky, path
length differences between station pairs are corrected, enabling the
formation of dynamic spectra in all four Stokes parameters at the
native time and frequency resolution. This step has been implemented
in \texttt{DynSpecMS}\footnote{\url{https://github.com/cyriltasse/DynSpecMS}}
software, and accounts for direction-dependent effects, such as
ionospheric and primary beam corrections, derived earlier in the
pipeline.

The burst of \stkm~ discussed here was detected in an $8$h scan, observed on 2016-05-17, and was located $\sim1.4$ degrees from the LoTSS pointing center named P236+53, having time and spectral resolution of 8\,s and 12.1\,kHz. Instrumental leakage between Stokes I and other Stokes parameters was measured to be
 $\lesssim1\%$ \citep{2023A&A...670A.124C}. This was further confirmed
 by analysing the dynamic spectra of other non-polarised sources in
 the field. The average rms noise in the Stokes\,V dynamic spectrum is $\approx$20\,mJy.

To produce Figure\,\ref{fig:circ_pol_image}, we used \texttt{WSClean} \citep{2014MNRAS.444..606O} with a robust parameter of $-0.5$ for only the data corresponding to the duration of the burst. Following the literature in this area \citep{callingham21,Callingham_crdra}, we define the sign of the Stokes V emission as left-hand circularly-polarised light minus right-hand circularly-polarised light.

\subsection*{X-ray Data Reduction}

On 2006-06-23, \emph{XMM-Newton} serendipitously observed \stkm~for $\approx$\,16\,ks at $\approx 13'$ away from the pointing centre (ObsID: 0401270201). We processed the observation following established routines \citep{Callingham2012,2019NatAs...3...82C}, utilising the \emph{XMM-Newton} Science Analysis System (SAS) version 20.0.0. Background flares were filtered from the European Photon Imaging Camera data during reduction. Spectral and timing data were extracted from a circular source region with a radius of approximately $16''$, while the background was sampled from a nearby region of double the area on the same CCD. Photon redistribution matrices and ancillary response files were generated for the spectral and timing analysis of the pn, MOS1, and MOS2 cameras using standard calibration files.

The average 0.2 to 10.0\,keV count rate for \stkm~during the observation was $\approx 1.1$ count\,s$^{-1}$, and no flares were detected during the observation. The spectrum of \stkm~was best fit by a single photoelectrically absorbed optically-thin plasma model (XSPEC \citep{Arnaud1996} model $\tt{apec}$), which
provides a 0.2 to 2.4\,keV flux of $(1.59\pm 0.05) \times$10$^{-12}$\,erg\,s$^{-1}$\,cm$^{-2}$. 

\subsection*{Coronal Scale Height and Plasma Emission}
The coronal temperature $T_{c}$ of \stkm~was estimated using the relation $T_{c} = 0.11F_{X}^{0.26}$, where $F_{X}$ represents the X-ray surface flux derived from the 0.2-2.4\,keV X-ray luminosity and the photospheric radius of the star \citep{2015A&A...578A.129J}. Therefore, the measured soft X-ray flux of $(1.59\pm 0.05) \times$10$^{-12}$\,erg\,s$^{-1}$\,cm$^{-2}$ implies a coronal temperature of $\sim$7.8\,MK. 

For estimating the shock velocity, the hydrostatic density structure of the corona of \stkm~was assumed to have a scale height \citep{2020NatAs.tmp...34V} of $H = 6 \times 10^{9} (T_{c}/10^{6}\,\mathrm{K})(R_{*}/R_{\odot})^{2}(M_{*}/M_{\odot})^{-1}$\,cm, where $R_{*}$ and $M_{*}$ are the stellar radius and mass, respectively. The uncertainty on the shock velocity is largely determined by the uncertainty on the drift, and the uncertainty in radius and mass of the star. Consistent with an M0V star, we used a radius of 0.63$\pm$0.02\,$R_{\odot}$ and mass of 0.61$\pm$0.01\,$M_{\odot}$ for the shock velocity calculation \citep{2019ApJ...871...63M}.

To calculate the brightness temperature of any potential plasma emission from \stkm, we varied the temperature of the injected hot plasma to be up to 40 times the measured coronal temperature and applied Equations 15 to 22 of \citep{Stepanov2001}. Such a temperature range is consistent, within an order of magnitude, with the observed temperatures of coronal loops \citep{Giampapa1996} and flares \citep{Robrade2010} on M~dwarfs. It is important to note that the hot component temperature being 40 times the ambient coronal density is chosen as the minimum temperature required to generate the observed brightness temperature, thus should not be the focus of physical interpretation. In particular, the level of turbulence in our plasma model was assumed to be $10^{-5}$, the canonical level at which the turbulence alters the dispersion in the background plasma for fundamental plasma emission \citep{Benz1993,Reid2017,Stepanov2001}. However, a different value for the level of turbulence could vastly change the required temperature of the injected hot component. The point of this exercise is to demonstrate that the properties of the radio emission can be self-consistently explained within a plasma emission model that does not require non-physical values.

Finally, we assumed the Alfv\'{e}n velocity $V_{a}$ in the stellar wind of \stkm~follows $V_{a} = B / (n_{p}m_{p}\mu_{0})^{0.5}$, where $B$, $n_{p}$, $m_{p}$, and $\mu_{0}$ are the magnetic field strength, number density of proton, mass of the proton, and magnetic permittivity of free space, respectively. For a Type\,II burst to occur, the velocity of the shock needs to exceed the Alfv\'{e}n velocity, implying $v_s >  B / (n_{p}m_{p}\mu_{0})^{0.5}$. Assuming quasi-neutrality ($n_{p} = n_{e}$), for $v_{s} = 2,400$\,km\,s$^{-1}$, the magnetic field strength of the CME at the location of emission must be less than 19\,G. Such a magnetic field strength is consistent with a dipole model (or solar model) that has a surface magnetic field strength of 300\,G and with the radio emission occuring at three stellar radii.

\subsection*{ECMI emission}

An alternative scenario to explain the time-frequency radio sweep observed from \stkm~is ECMI emission originating in a flaring loop on the stellar surface. ECMI emission is characterised by its beam pattern, which resembles a hollow cone centered on the magnetic field line at the emission site. Due to the anisotropic beam pattern, an observer will generally observe ECMI emission to manifest in short bursts as the emission cone sweeps across the line of sight \citep{Louis2019, Kavanagh2023, kavanagh24}. It also occurs at the local cyclotron frequency, which scales linearly with the local magnetic field strength $B$ \citep{1985ARA&A..23..169D}
\begin{equation}
\nu = 2.8 \times B~\text{MHz},
\label{eq:cyclotron frequency}
\end{equation}
where $B$ is in Gauss. Therefore, if ECMI emission occurs along a flaring loop, one can contrive for certain geometric configurations of the viewing angle of the loop from the observer's perspective to produce a time-frequency sweep akin to that seen from \stkm~as the star rotates \citep[e.g.][]{Louis2019}.

Recently, radio emission from flaring loops on the Sun suggestive of ECMI has been observed \cite{yu24}, albeit without any significant drift in time, which is likely due to the longer rotation period of the Sun compared to \stkm. To test the viability of the flaring loop scenario for \stkm, we construct a simple geometric model. We place a circular loop of radius $L$ centered at the stellar surface \citep[e.g.][]{titov99}. The co-latitude of the loop center is $\theta_l$, and its longitude at some reference time $t_0$ is $\phi_l$. The loop is misaligned with the meridian by the angle $\delta_l$. The magnetic field strength at the footpoints where the loop meets the surface is $B_\text{fp}$. Spatially-resolved observations of ECMI-like emission from flaring loops on the the Sun indicate that the magnetic field strength along the loop drops off roughly linearly with the distance above the surface \citep{yu24}. We therefore adopt a linear prescription for the magnetic field strength in the loop, which drops off linearly with the rate $m$
\begin{equation}
B_l(r) = B_\text{fp} - m (r - R_\star),
\label{eq:flaring loop field strength}
\end{equation}
where $r$ is the radial distance between the points on the loop and the centre of the star. 

We assume that each point on the flaring loop emits ECMI emission, which is beamed outward in a cone shape. The emission cones are aligned with the magnetic field vector at each point on the loop, each of which being tangent to the loop. The cone opens out from the magnetic field vector by the $\alpha$, and has a thickness $\Delta\alpha$. We calculate the angle $\gamma$ between the vector at each emission site and the line of sight, and use it to prescribe the flux density visible to the observer \citep{kavanagh24}
\begin{equation}
F = F_0\exp\Big[-\frac{1}{2} \Big(\frac{\gamma - \alpha}{\Delta\alpha}\Big)^2\Big] ,
\end{equation}
where $F_0$ is the maximum flux density visible. In other words, the bulk of the flux density seen when $\gamma$ is within the range of $\sim\alpha\pm\Delta\alpha / 2$. The frequency bins of the dynamic spectrum provide the emission frequency for each point on the loop via Equations~\ref{eq:cyclotron frequency} and \ref{eq:flaring loop field strength}, and the time bins provide the longitude of the loop $\phi$ as a function of time $t$
\begin{equation}
\phi = \phi_l + \frac{2\pi}{P} (t - t_0),
\end{equation}
where $P = 1.24$~days is the rotation period. 

To explore the likelihood space of the model parameters given the burst shown in Figure~\ref{fig:dyn_spec_all}, we use UltraNest\footnote{\url{https://johannesbuchner.github.io/UltraNest/}} \citep{buchner21a}, which utilises the nested sampling Monte Carlo algorithm MLFriends \citep{buchner16, buchner19}. Uniform priors are assumed for each parameter, which are listed in Extended Data Table~\ref{table:ECM params}.

Due to the brightness of the burst, the noise in the dynamic spectrum is not scattered around zero, and instead sits at a value of $\sim20$~mJy. To avoid UltraNest attempting to fit this component of the dynamic spectrum, we subtract this value from each pixel prior to running the sampler. The noise of each pixel is $\sim20$~mJy. Otherwise, our setup of UltraNest is identical to the approach taken in \cite{kavanagh24}.

Within 50 steps, the sampler converges on the result shown in Figure~\ref{fig:dyn spec ECM model}, with a reduced $\chi^2$ of 1.95. While we reproduce the overall drifting structure observed from \stkm, we cannot recover the finer sub-structure seen in Extended Data Figure\,1. The confidence intervals inferred for each model parameter are listed in Extended Data Table~\ref{table:ECM params}.

We infer a flaring loop size of 0.3~$R_\star$, which is comparable to that obtained on the Sun by \cite{yu24}, which we visually estimate from Figure 4 of \cite{yu24} to be $\sim0.2~R_\odot$. Our magnetic field gradient of 9583 G~${R_\star}^{-1}$ is also comparable to \cite{yu24}, who obtain a value of 2951 G~${R_\odot}^{-1}$ (again estimated visually from their Figure 4). The obtained cone opening angle is also in agreement with observations of ECMI on Jupiter \citep{Hess2008}.

However, we note the only way to get this model to fit is to have a incredibly thin cone thickness of 0.01\degr, which is significantly smaller than the $\sim1$\degr\ inferred from observations of ECMI on Jupiter  \citep{kaiser00}. While ECMI-like radio bursts have recently observed on the Sun that could also have such thin cone thickness, such busts only last for about 10 seconds \citep[see Extended Data Figure 8 of][]{yu24}.

If the observed burst reported here for \stkm~and on the Sun by \cite{yu24} are indeed driven by ECMI, follow-up studies focusing on the theoretical feasibility for producing extremely narrow ECMI will be needed to confirm this scenario. The main issue is that the electron velocity distribution is incredibly small and not predicted by standard theory \citep[e.g.][]{Treumann2006}.

Another uncertainty about our inferred parameters is the high latitude (or low co-latitude) of the flaring loop. While high-latitude flares on M dwarfs have also been observed \citep[e.g.][]{ilin21}, it remains unclear if this is standard for all M dwarfs since pervious studies have been restricted to active, rapidly rotating ($<$0.5\,d) M dwarfs. We also did not re-observe the star in radio despite observing it with LOFAR for subsequent rotations. Radio emission driven by ECMI on M dwarfs have been observed over multiple rotations \citep{Callingham_crdra}.

\bmhead{Acknowledgments}

This manuscript is dedicated to the memory of Judy Ann Callingham, in recognition of her lifelong support of the first author. 

We thank the referees (R. Osten and two anonymous) for their useful comments. JRC acknowledges funding from the European Union via the European Research Council (ERC) grant Epaphus (project number 101166008). RDK and HKV acknowledge funding from the Dutch Research Council (NWO) for the project `e-MAPS' (project number Vi.Vidi.203.093) under the NWO talent scheme VIDI. SB acknowledges funding by the NWO under the project ``Exo-space weather and contemporaneous signatures of star-planet interactions" (with project number OCENW.M.22.215 of the research programme ``Open Competition Domain Science- M"). PZarka  acknowledges funding from the ERC under the European Union’s Horizon 2020 research and innovation programme (grant agreement no. 101020459 - Exoradio). HR acknowledges UKSA grant ST/X002012/1 and STFC grant ST/W001004/1.

The LOFAR data in this manuscript were (partly) processed by the LOFAR Two-Metre Sky Survey (LoTSS) team. This team made use of the LOFAR direction independent calibration pipeline (\url{https://github.com/lofar-astron/prefactor}), which was deployed by the LOFAR e-infragroup on the Dutch National Grid infrastructure with support of the SURF Co-operative through grants e-infra 170194 e-infra 180169. The LoTSS direction dependent calibration and imaging pipeline (\url{http://github.com/mhardcastle/ddf-pipeline/}) was run on compute clusters at Leiden Observatory and the University of Hertfordshire, which are supported by a European Research Council Advanced Grant [NEWCLUSTERS-321271] and the UK Science and Technology Funding Council [ST/P000096/1]. The J\"ulich LOFAR Long Term Archive and the German LOFAR network are both coordinated and operated by the J\"ulich Supercomputing Centre (JSC), and computing resources on the supercomputer JUWELS at JSC were provided by the Gauss Centre for Supercomputing e.V. (grant CHTB00) through the John von Neumann Institute for Computing (NIC).

\section*{Declarations}

\begin{itemize}
\item Conflict of interest: The authors do not have any conflicts of interest to report.

\item Authors' contributions: JRC initiated the LOFAR project that led to the discovery of the source, conducted the cross-matching analysis and wrote the paper. CT wrote the dynamic spectra software and helped interpret the data. RK identified the burst. HKV, JRC, RDK and PZarka led the theoretical interpretation of the detection and contributed substantially to the paper. SBellotti and PIC obtained the Zeeman Doppler imaging data on the star. MJH and TWS processed the survey data. PZucca and HR provided solar physics expertise for interpreting the burst. SBloot, DCK, LL, EKP and BJSP commented on the paper and provided relevant expertise in characterizing StKM 1-1262. HJAR is the principal investigator of LoTSS and commented on the paper.

\item Data Availability: LOFAR visibilities taken are publicly available via the LOFAR Long Term Archive (ObsID: 470106; LoTSS field: P236+53). The \emph{XMM-Newton} data are available through the \emph{XMM-Newton} Science Archive (XSA) (ObsID: 0401270201). All other data used in the manuscript have been sourced from the public domain.

\item Code Availability: The important codes used to analyse and process the LOFAR data are available at the following sites:
    WSClean (\url{https://gitlab.com/aroffringa/wsclean}),
    DynSpecMS (\url{https://github.com/cyriltasse/DynSpecMS}),
    and DDF pipeline (\url{https://github.com/mhardcastle/ddf-pipeline}). The posteriors obtained from fitting a geometric flare model to the dynamic radio spectrum are available here: \url{https://github.com/robkavanagh/papers/tree/main/type-II}.
    
\item Correspondence: Correspondence and requests for materials should be addressed to JRC and CT \\(email: callingham@astron.nl and cyril.tasse@obspm.fr).
\item Inclusion \& Ethics statement: All relevant ethics and inclusion principles for an astronomy project using data from LOFAR have been followed.
\end{itemize}

\clearpage

\section*{Extended Data}

\begin{figure}[h]
\setcounter{figure}{0}
\renewcommand\figurename{Extended Data Figure}
\centering
\includegraphics[width=0.7\linewidth]{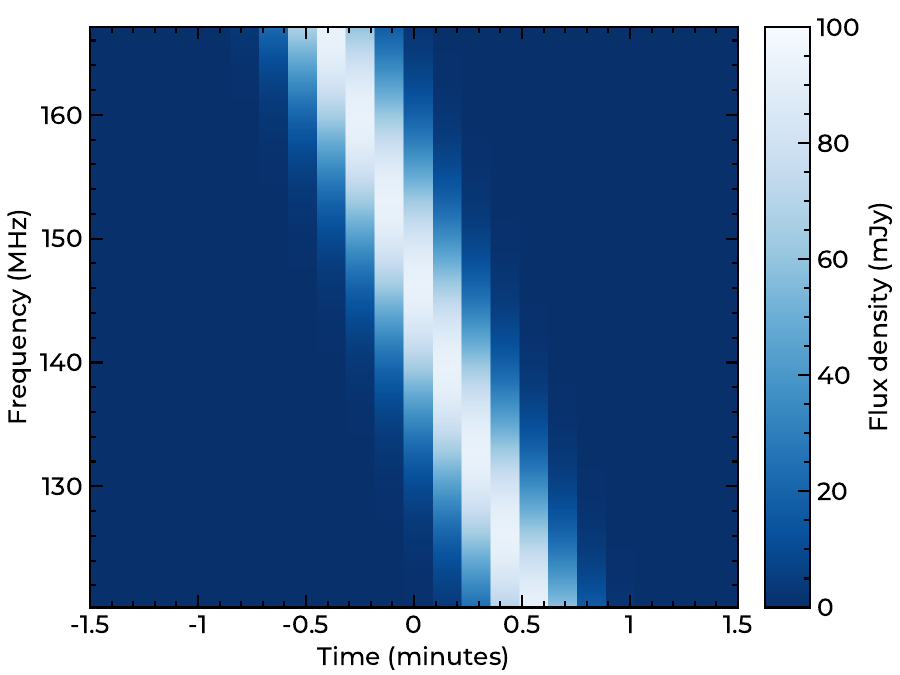}
\caption{\textbf{Reconstruction of the burst assuming ECMI from a high-latitude coronal loop.} While the overall drift rate is recovered, it is not possible to recover the observed sub-structure in Figure\,\ref{fig:dyn_spec_all}.}
\label{fig:dyn spec ECM model}
\end{figure}

\newpage 

\begin{table}[h]
\renewcommand\tablename{Extended Table}
\caption{Fit parameters for the ECMI emission model. The uniform prior ranges and posteriors obtained for each parameter in our flaring loop model. The Julian Date for $t_0$ is 2457526.328. Note that the uncertainty on $\Delta\alpha$ is smaller than the lowest significant digit. The uncertainties presented in this table represent 1-$\sigma$.}
\label{table:ECM params}
\centering
\begin{tabular}{lccc}
Parameter & Symbol & Prior & Posterior \\
\hline
Loop co-latitude (\degr) & $\theta_l$ & [0, 90] & $16.65_{-0.47}^{+1.36}$ \\
Loop longitude at time $t_0$ (\degr) & $\phi_l$ & [-90, 90] & $-73.30_{-4.15}^{+2.19}$ \\
Angle between loop and the meridion (\degr) & $\delta_l$ & [0, 180] & $120.08_{-3.30}^{+1.45}$ \\
Magnetic field strength at footpoint (G) & $B_\text{fp}$ & [60, 10000] & $174.29_{-40.88}^{+45.53}$ \\
Magnetic field gradient (G ${R_\star}^{-1}$) & $m$ & [0, 20000] & $9583.77_{-754.81}^{+491.12}$ \\
Loop radius ($R_\star$) & $L$ & [0, 0.5] & $0.30\pm0.02$ \\
Cone opening angle (\degr) & $\alpha$ & [40, 90] & $77.42_{-1.83}^{+0.88}$ \\
Cone thickness (\degr) & $\Delta\alpha$ & [0, 1] & $0.01$ \\
Maximum flux density visible (mJy) & $F_0$ & [0, 500] & $92.84_{-1.90}^{+1.85}$ \\
\end{tabular}
\end{table}

\newpage







\end{document}